\documentclass[preprint,superscriptaddress,nofootinbib]{revtex4}
  \usepackage{graphicx}
  \begin{document}
  \title{The ${\Upsilon}(1S)$ ${\to}$ $B_{c}{\rho}$ decay with perturbative QCD approach}
  \author{Junfeng Sun}
  \affiliation{Institute of Particle and Nuclear Physics,
              Henan Normal University, Xinxiang 453007, China}
  \author{Yueling Yang}
  \affiliation{Institute of Particle and Nuclear Physics,
              Henan Normal University, Xinxiang 453007, China}
  \author{Qingxia Li}
  \affiliation{Institute of Particle and Nuclear Physics,
              Henan Normal University, Xinxiang 453007, China}
  \author{Gongru Lu}
  \affiliation{Institute of Particle and Nuclear Physics,
              Henan Normal University, Xinxiang 453007, China}
  \author{Jinshu Huang}
  \affiliation{College of Physics and Electronic Engineering,
              Nanyang Normal University, Nanyang 473061, China}
  \author{Qin Chang}
  \affiliation{Institute of Particle and Nuclear Physics,
              Henan Normal University, Xinxiang 453007, China}

  \begin{abstract}
  With the potential prospects of the ${\Upsilon}(1S)$ data
  samples at the running LHC and upcoming SuperKEKB,
  the ${\Upsilon}(1S)$ ${\to}$ $B_{c}{\rho}$
  weak decay is studied with the pQCD approach.
  It is found that (1) the lion's share of branching ratio
  comes from the longitudinal polarization helicity
  amplitudes; (2) branching
  ratio for the ${\Upsilon}(1S)$ ${\to}$ $B_{c}{\rho}$
  decay can reach up to ${\cal O}(10^{-9})$, which might
  be hopefully measurable.
  \end{abstract}
  \maketitle

  \section{Introduction}
  \label{sec01}
  The ${\Upsilon}(1S)$ meson consists of the bottom quark
  and antiquark pair $b\bar{b}$, carries the definitely
  established quantum numbers of $I^{G}J^{PC}$ $=$
  $0^{-}1^{--}$ \cite{pdg}, and lies below the kinematic
  $B\bar{B}$ threshold.
  The ${\Upsilon}(1S)$ meson decay mainly through the strong
  interaction, the electromagnetic interaction and radiative
  transition. Besides, the ${\Upsilon}(1S)$ meson can also
  decay via the weak interactions within the standard model.
  More than $10^{8}$ ${\Upsilon}(1S)$ data samples have been
  accumulated at Belle \cite{epjc74}. More and more
  upsilon data samples with high precision are promisingly
  expected at the running LHC and the forthcoming SuperKEKB.
  Although the branching ratio for the ${\Upsilon}(1S)$ weak
  decay is tiny, it seems to exist a realistic possibility
  to search for the signals of the ${\Upsilon}(1S)$ weak
  decay at future experiments.
  In this paper, we will study the ${\Upsilon}(1S)$ ${\to}$
  $B_{c}{\rho}$ weak decay with the
  perturbative QCD (pQCD) approach \cite{pqcd1,pqcd2,pqcd3}.

  Experimentally, there is no report on the ${\Upsilon}(1S)$ ${\to}$
  $B_{c}{\rho}$ weak decay so far.
  The signals for the ${\Upsilon}(1S)$ ${\to}$ $B_{c}{\rho}$ weak
  decay should, in principle, be easily identified, due to the facts
  that the final states have different electric charges, have definite
  momentum and energy, and are back-to-back in the rest frame of the
  ${\Upsilon}(1S)$ meson.
  In addition, the identification of a single flavored $B_{c}$ meson
  could be used to effectively enhance signal-to-background ratio.
  Another important and fashionable motivation is that evidences
  of an abnormally large branching ratio for the
  ${\Upsilon}(1S)$ weak decay might be a hint of new physics.

  Theoretically, the ${\Upsilon}(1S)$ ${\to}$ $B_{c}{\rho}$ weak decay
  belongs to the external $W$ emission topography, and is favored by
  the Cabibbo-Kabayashi-Maskawa (CKM) matrix elements
  ${\vert}V_{cb}V_{ud}^{\ast}{\vert}$.
  So it should have relatively large branching ratio
  among the ${\Upsilon}(1S)$ weak decays, which has been
  studied with the naive factorization (NF) approximation
  \cite{ijma14,adv2013}.
  Recently, some attractive methods have been
  developed, such as the pQCD approach \cite{pqcd1,pqcd2,pqcd3},
  the QCD factorization approach \cite{qcdf1,qcdf2,qcdf3},
  soft and collinear effective theory \cite{scet1,scet2,scet3,scet4},
  and applied widely to accommodate measurements on
  the $B$ meson weak decays.
  The ${\Upsilon}(1S)$ ${\to}$ $B_{c}{\rho}$ decay
  permit one to cross check parameters obtained from the
  $B$ meson decay, to test the practical applicability
  of various phenomenological models in the vector meson
  weak decays, and to further explore the underlying dynamical
  mechanism of the heavy quark weak decay.
  In addition, as it is well known, the $B_{c}$ meson carries
  two explicit heavy flavors and has extremely abundant
  decay modes, but its hadronic production is suppressed
  compared with that for hidden-flavor quarkonia and
  heavy-light mesons, due to higher order in QCD coupling
  constants ${\alpha}_{s}$ and the presence of additional
  heavy quarks \cite{bc1,bc2}.
  The ${\Upsilon}(1S)$ ${\to}$ $B_{c}{\rho}$ decay
  offers another platform to study the $B_{c}$ meson production
  at high energy colliders.

  This paper is organized as follows.
  In section \ref{sec02}, we present the theoretical framework
  and the amplitudes for the ${\Upsilon}(1S)$ ${\to}$ $B_{c}{\rho}$
  decay with the pQCD approach.
  Section \ref{sec03} is devoted to numerical results and discussion.
  The last section is our summary.

  \section{theoretical framework}
  \label{sec02}
  \subsection{The effective Hamiltonian}
  \label{sec0201}
  The effective Hamiltonian responsible for the
  ${\Upsilon}(1S)$ ${\to}$ $B_{c}{\rho}$ weak decay is
  \cite{9512380}
   \begin{equation}
  {\cal H}_{\rm eff}\ =\ \frac{G_{F}}{\sqrt{2}}\,
   V_{cb} V_{ud}^{\ast}\,
   \Big\{ C_{1}({\mu})\,Q_{1}({\mu})
         +C_{2}({\mu})\,Q_{2}({\mu}) \Big\}
   + {\rm H.c.}
   \label{hamilton},
   \end{equation}
  where $G_{F}$ ${\simeq}$ $1.166{\times}10^{-5}\,{\rm GeV}^{-2}$ \cite{pdg}
  is the Fermi coupling constant; the CKM factor is written
  as a power series in the Wolfenstein parameter ${\lambda}$
  ${\simeq}$ $0.2$ \cite{pdg},
  \begin{equation}
  V_{cb}V_{ud}^{\ast}\ =\
               A{\lambda}^{2}
  - \frac{1}{2}A{\lambda}^{4}
  - \frac{1}{8}A{\lambda}^{6}
  +{\cal O}({\lambda}^{8})
  \label{eq:ckm01}.
  \end{equation}

  The local operators are defined as follows:
    \begin{eqnarray}
    Q_{1} &=&
  [ \bar{c}_{\alpha}{\gamma}_{\mu}(1-{\gamma}_{5})b_{\alpha} ]
  [ \bar{q}_{\beta} {\gamma}^{\mu}(1-{\gamma}_{5})u_{\beta} ]
    \label{q1}, \\
    Q_{2} &=&
  [ \bar{c}_{\alpha}{\gamma}_{\mu}(1-{\gamma}_{5})b_{\beta} ]
  [ \bar{q}_{\beta}{\gamma}^{\mu}(1-{\gamma}_{5})u_{\alpha} ]
    \label{q2},
    \end{eqnarray}
  where ${\alpha}$ and ${\beta}$ are color indices.

  From Eq.(\ref{hamilton}), it is clearly seen that
  only the tree operators contribute to the concerned
  process, and there is no pollution from penguin and
  annihilation contributions.
  As it is well known, degrees of freedom with mass scales
  above ${\mu}$ are integrated out into the
  Wilson coefficients $C_{1,2}(\mu)$ typically using
  the renormalization group assisted perturbation
  theory. The physical contributions below the scale
  of ${\mu}$ are included in the hadronic matrix
  elements (HME) where the local operators sandwiched
  between initial and final hadron states.
  The most complicated part is the treatment on HME,
  where the perturbative and nonperturbative
  effects entangle with each other.
  To obtain the decay amplitudes, the remaining work
  is to calculate HME properly.

  \subsection{Hadronic matrix elements}
  \label{sec0202}
  With the Lepage-Brodsky approach for exclusive processes \cite{prd22},
  HME could be expressed as the convolution of hard scattering
  subamplitudes containing perturbative contributions with the
  universal wave functions reflecting the nonperturbative
  contributions. To eliminate the endpoint singularities appearing
  in the collinear factorization approximation, the pQCD approach
  suggests \cite{pqcd1,pqcd2,pqcd3} retaining the transverse
  momentum of quarks and introducing the Sudakov factor.
  Finally,
  the decay amplitudes could be factorized into three parts
  \cite{pqcd2,pqcd3}: the hard effects enclosed by the Wilson coefficients
  $C_{i}$, the heavy quark decay subamplitudes ${\cal H}$, and the
  universal wave functions ${\Phi}$,
  \begin{equation}
  {\int} dk\,
  C_{i}(t)\,{\cal H}(t,k)\,{\Phi}(k)\,e^{-S}
  \label{hadronic},
  \end{equation}
  where $t$ is a typical scale, $k$ is the momentum of the valence
  quarks, and the Sudakov factor $e^{-S}$ can effectively suppress
  the long-distance contributions and make the hard
  scattering more perturbative.

  \subsection{Kinematic variables}
  \label{sec0203}
\begin{figure}[h]
  \includegraphics[width=0.99\textwidth,bb=75 620 530 725]{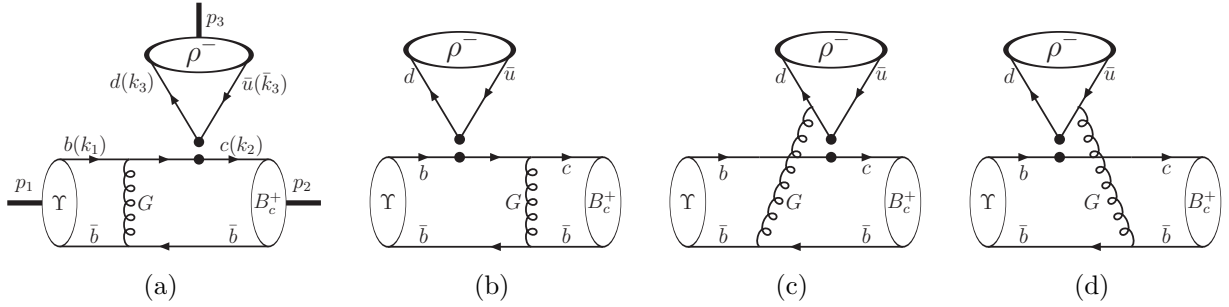}
  \caption{Feynman diagrams for the ${\Upsilon}$ ${\to}$ $B_{c}{\rho}$
   decay with the pQCD approach, where (a) and (b) are factorizable
   emission diagrams, (c) and (d) are nonfactorizable emission
   diagrams.}
  \label{feynman}
  \end{figure}

  The light cone kinematic variables in the ${\Upsilon}(1S)$
  rest frame are defined as follows:
  \begin{equation}
  p_{{\Upsilon}}\, =\, p_{1}\, =\, \frac{m_{1}}{\sqrt{2}}(1,1,0)
  \label{kine-p1},
  \end{equation}
  \begin{equation}
  p_{B_{c}}\, =\, p_{2}\, =\, (p_{2}^{+},p_{2}^{-},0)
  \label{kine-p2},
  \end{equation}
  \begin{equation}
  p_{\rho}\, =\, p_{3}\, =\, (p_{3}^{-},p_{3}^{+},0)
  \label{kine-p3},
  \end{equation}
  \begin{equation}
  k_{i}\, =\, x_{i}\,p_{i}+(0,0,\vec{k}_{i{\perp}})
  \label{kine-ki},
  \end{equation}
  \begin{equation}
 {\epsilon}_{i}^{\parallel}\, =\,
  \frac{p_{i}}{m_{i}}-\frac{m_{i}}{p_{i}{\cdot}n_{+}}n_{+}
  \label{kine-longe},
  \end{equation}
  \begin{equation}
 {\epsilon}_{i}^{\perp}\, =\, (0,0,\vec{1})
  \label{kine-transe},
  \end{equation}
  \begin{equation}
  n_{+}=(1,0,0)
  \label{kine-null},
  \end{equation}
  \begin{equation}
  p_{i}^{\pm}\, =\, (E_{i}\,{\pm}\,p)/\sqrt{2}
  \label{kine-pipm},
  \end{equation}
  \begin{equation}
  s\, =\, 2\,p_{2}{\cdot}p_{3}
  \label{kine-s},
  \end{equation}
  \begin{equation}
  t\, =\, 2\,p_{1}{\cdot}p_{2}\, =\ 2\,m_{1}\,E_{2}
  \label{kine-t},
  \end{equation}
  \begin{equation}
  u\, =\, 2\,p_{1}{\cdot}p_{3}\, =\ 2\,m_{1}\,E_{3}
  \label{kine-u},
  \end{equation}
  \begin{equation}
  p = \frac{\sqrt{ [m_{1}^{2}-(m_{2}+m_{3})^{2}]\,[m_{1}^{2}-(m_{2}-m_{3})^{2}] }}{2\,m_{1}}
  \label{kine-pcm},
  \end{equation}
  where $x_{i}$ and $\vec{k}_{i{\perp}}$ are the longitudinal momentum
  fraction and transverse momentum of the valence quark, respectively;
  ${\epsilon}_{i}^{\parallel}$ and ${\epsilon}_{i}^{\perp}$ are the
  longitudinal and transverse polarization vectors, respectively,
  satisfying with the relations ${\epsilon}_{i}^{2}$ $=$ $-1$
  and ${\epsilon}_{i}{\cdot}p_{i}$ $=$ $0$;
  the subscript $i$ $=$ $1$, $2$, $3$ on variables ($p_{i}$, $E_{i}$, $m_{i}$
  and ${\epsilon}_{i}^{{\parallel},{\perp}}$) correspond to the
  ${\Upsilon}(1S)$, $B_{c}$ and ${\rho}$ mesons, respectively;
  $n_{+}$ is the null vector; $s$, $t$ and $u$ are the Lorentz-invariant
  variables; $p$ is the common momentum of final states.
  The notation of momentum is displayed in Fig.\ref{feynman}(a).

  \subsection{Wave functions}
  \label{sec0204}
  With the notation in \cite{prd65,jhep0605}, the definitions of
  the diquark operator HME are
  \begin{equation}
 {\langle}0{\vert}b_{i}(z)\bar{b}_{j}(0){\vert}
 {\Upsilon}(p_{1},{\epsilon}_{1}^{\parallel}){\rangle}\,
 =\, \frac{f_{\Upsilon}}{4}{\int}d^{4}k_{1}\,e^{-ik_{1}{\cdot}z}
  \Big\{ \!\!\not{\epsilon}_{1}^{{\parallel}} \Big[
   m_{1}\,{\Phi}_{\Upsilon}^{v}(k_{1})
  -\!\!\not{p}_{1}\, {\Phi}_{\Upsilon}^{t}(k_{1})
  \Big] \Big\}_{ji}
  \label{wave-bbl},
  \end{equation}
  \begin{equation}
 {\langle}0{\vert}b_{i}(z)\bar{b}_{j}(0){\vert}
 {\Upsilon}(p_{1},{\epsilon}_{1}^{{\perp}}){\rangle}\,
 =\, \frac{f_{\Upsilon}}{4}{\int}d^{4}k_{1}\,e^{-ik_{1}{\cdot}z}
  \Big\{ \!\!\not{\epsilon}_{1}^{{\perp}} \Big[
   m_{1}\,{\Phi}_{\Upsilon}^{V}(k_{1})
  -\!\!\not{p}_{1}\, {\Phi}_{\Upsilon}^{T}(k_{1})
  \Big] \Big\}_{ji}
  \label{wave-bbt},
  \end{equation}
  \begin{equation}
 {\langle}B_{c}(p_{2}){\vert}\bar{c}_{i}(z)b_{j}(0){\vert}0{\rangle}\,
 =\, \frac{i}{4}f_{B_{c}} {\int}dx_{2}\,e^{ix_{2}p_{2}{\cdot}z}\,
  \Big\{ {\gamma}_{5}\Big[ \!\!\not{p}_{2}+m_{2}\Big]
 {\phi}_{B_{c}}(x_{2}) \Big\}_{ji}
  \label{wave-bcp},
  \end{equation}
  \begin{equation}
 {\langle}{\rho}(p_{3},{\epsilon}_{3}^{{\parallel}})
 {\vert}u_{i}(0)\bar{d}_{j}(z){\vert}0{\rangle}\, =\,
  \frac{1}{4}{\int}_{0}^{1}dk_{3}\,e^{ik_{3}{\cdot}z}
  \Big\{ \!\!\not{\epsilon}_{3}^{{\parallel}}\,
   m_{3}\,{\Phi}_{\rho}^{v}(k_{3})
  +\!\!\not{\epsilon}_{3}^{{\parallel}}
   \!\!\not{p}_{3}\, {\Phi}_{\rho}^{t}(k_{3})
  +m_{3}\,{\Phi}_{\rho}^{s}(k_{3}) \Big\}_{ji}
  \label{wave-rholong},
  \end{equation}
  \begin{eqnarray}
  \lefteqn{
 {\langle}{\rho}(p_{3},{\epsilon}_{3}^{{\perp}})
 {\vert}u_{i}(0)\bar{d}_{j}(z){\vert}0{\rangle}\ =\
  \frac{1}{4}{\int}_{0}^{1}dk_{3}\,e^{ik_{3}{\cdot}z}
  \Big\{ \!\not{\epsilon}_{3}^{{\perp}}
   m_{3}\,{\Phi}_{\rho}^{V}(k_{3})
  }
  \nonumber \\ & &
  +\!\not{\epsilon}_{3}^{{\perp}}
   \!\not{p}_{3}\, {\Phi}_{\rho}^{T}(k_{3})
  +\frac{i\,m_{3}}{p_{3}{\cdot}n_{+}}
  {\varepsilon}_{{\mu}{\nu}{\alpha}{\beta}}\,
  {\gamma}_{5}\,{\gamma}^{\mu}\,{\epsilon}_{3}^{{\perp},{\nu}}\,
  p_{3}^{\alpha}\,n_{+}^{\beta}\,
  {\Phi}_{\rho}^{A}(k_{3}) \Big\}_{ji}
  \label{wave-rhotrans},
  \end{eqnarray}
  where $f_{\Upsilon}$ and $f_{B_{c}}$ are decay constants;
  the definitions of wave functions ${\Phi}_{\rho}^{v,t,s}$
  and ${\Phi}_{\rho}^{V,T,A}$ can be found in Ref. \cite{prd65,jhep0605}.
  In fact, for the ${\rho}$ meson, only three wave functions
  ${\Phi}_{\rho}^{v}$ and ${\Phi}_{\rho}^{V,A}$ are involved
  in the decay amplitudes (see Appendix \ref{blocks}).
  The twist-2 distribution amplitude for the longitudinal
  polarization ${\rho}$ meson is \cite{prd65,jhep0605}:
  \begin{equation}
 {\phi}_{\rho}^{v}(x) \, =\, f_{\rho}\,
  6\,x\,\bar{x} \sum\limits_{i=0} a^{\parallel}_{2i}\,C_{2i}^{3/2}(t)
  \label{wave-rhov},
  \end{equation}
  where $f_{\rho}$ is the decay constant;
  $\bar{x}$ $=$ $1$ $-$ $x$; $t$ $=$ $\bar{x}$ $-$ $x$;
  $a^{\parallel}_{i}$ and $C_{i}^{3/2}(t)$ are the Gegenbauer moment
  and polynomial, respectively; $a^{\parallel}_{i}$ $=$ $0$ for
  odd $i$ due to the $G$-parity invariance of the ${\rho}$
  distribution amplitudes.
  As to the twist-3 distribution amplitudes of the transverse
  polarization ${\rho}$ meson, for simplicity, we will take
  their asymptotic forms \cite{prd65,jhep0605}:
  \begin{equation}
 {\phi}_{\rho}^{V}(x) \, =\, f_{\rho}\, \frac{3}{4}\, (1+t^{2})
  \label{wave-rhoV},
  \end{equation}
  \begin{equation}
 {\phi}_{\rho}^{A}(x) \, =\, f_{\rho}\, \frac{3}{2}\, (-t)
  \label{wave-rhoA}.
  \end{equation}

  Because of $m_{{\Upsilon}(1S)}$ ${\simeq}$ $2m_{b}$
  and $m_{B_{c}}$ ${\simeq}$ $m_{b}$ $+$ $m_{c}$,
  both ${\Upsilon}(1S)$ and $B_{c}$ systems are nearly
  nonrelativistic.
  Nonrelativistic quantum chromodynamics (NRQCD)
  \cite{prd46,prd51,rmp77} and Schr\"{o}dinger
  equation can be used to describe their spectrum.
  The eigenfunction of the time-independent
  Schr\"{o}dinger equation with scalar harmonic
  oscillator potential corresponding to the quantum
  numbers $nL$ $=$ $1S$ is written as
   \begin{equation}
  {\phi}(\vec{k})\
  {\sim}\ e^{-\vec{k}^{2}/2{\beta}^{2}}
   \label{wave-k},
   \end{equation}
  where parameter ${\beta}$ determines the average transverse
  momentum, i.e.,
  ${\langle}1S{\vert}\vec{k}^{2}_{\perp}{\vert}1S{\rangle}$
  $=$ ${\beta}^{2}$.
  Employing the Brodsky-Huang-Lepage ansatz \cite{huang,xiao}
  which has been used to structure wave functions for light
  and heavy mesons \cite{wu2},
   \begin{equation}
   \vec{k}^{2}\ {\to}\ \frac{1}{4} \sum\limits_{i}
   \frac{\vec{k}_{i\perp}^{2}+m_{q_{i}}^{2}}{x_{i}}
   \label{wave-kt},
   \end{equation}
  where $x_{i}$, $\vec{k}_{i\perp}$, $m_{q_{i}}$ are the
  longitudinal momentum fraction, transverse momentum,
  mass of the valence quarks in hadrons, respectively,
  with the relations ${\sum}x_{i}$ $=$ $1$ and
  $\sum\vec{k}_{i\perp}$ $=$ $0$, then
  integrating out $\vec{k}_{i\perp}$ and combining with
  their asymptotic forms, one can obtain \cite{prd65,fbb}
   \begin{equation}
  {\phi}_{B_{c}}(x) = A\, x\bar{x}\,
  {\exp}\Big\{ -\frac{\bar{x}\,m_{c}^{2}+x\,m_{b}^{2}}
                     {8\,{\beta}_{2}^{2}\,x\,\bar{x}} \Big\}
   \label{wave-bc},
   \end{equation}
   \begin{equation}
  {\phi}_{\Upsilon}^{v}(x) = {\phi}_{\Upsilon}^{T}(x) = B\, x\bar{x}\,
  {\exp}\Big\{ -\frac{m_{b}^{2}}{8\,{\beta}_{1}^{2}\,x\,\bar{x}} \Big\}
   \label{wave-bbv},
   \end{equation}
   \begin{equation}
  {\phi}_{\Upsilon}^{t}(x) = C\, t^{2}\,
  {\exp}\Big\{ -\frac{m_{b}^{2}}{8\,{\beta}_{1}^{2}\,x\,\bar{x}} \Big\}
   \label{wave-bblt},
   \end{equation}
   \begin{equation}
  {\phi}_{\Upsilon}^{V}(x) = D\, (1+t^{2})\,
  {\exp}\Big\{ -\frac{m_{b}^{2}}{8\,{\beta}_{1}^{2}\,x\,\bar{x}} \Big\}
   \label{wave-bbtt},
   \end{equation}
   where the exponential function represents the transverse
   momentum distribution and can suppress the end-point
   singularity;
   ${\beta}_{i}$ ${\simeq}$ ${\xi}_{i}{\alpha}_{s}({\xi}_{i})$
   with ${\xi}_{i}$ $=$ $m_{i}/2$ based on the NRQCD power
   counting rules \cite{prd46};
   parameters $A$, $B$, $C$, $D$ are the normalization coefficients
   satisfying the conditions
   \begin{equation}
  {\int}_{0}^{1}dx\,{\phi}_{B_{c}}(x)=1,
   \quad
  {\int}_{0}^{1}dx\,{\phi}_{\Upsilon}^{v,t}(x) =
  {\int}_{0}^{1}dx\,{\phi}_{\Upsilon}^{V,T}(x) =1
   \label{wave-abc}.
   \end{equation}

  The shape lines for the normalized distribution amplitudes
  of ${\phi}_{B_{c}}(x)$ and ${\phi}_{\Upsilon}^{v,t,V,T}(x)$
  have been displayed in Fig.1 of Ref.\cite{sun16}, from
  which one can see that Eqs.(\ref{wave-bc})-(\ref{wave-bbtt})
  reflect generally the feature that valence quarks of hadrons
  share momentum fractions according to their masses.

  \subsection{Decay amplitudes}
  \label{sec0205}
  The Feynman diagrams for the ${\Upsilon}(1S)$ ${\to}$
  $B_{c}{\rho}$ decay are shown in Fig.\ref{feynman},
  including factorizable emission topologies (a) and (b)
  where gluon connects to the quarks in the same meson,
  and nonfactorizable emission topologies (c) and (d)
  where gluon attaches to the quarks in two different mesons.

  The amplitude for the ${\Upsilon}(1S)$ ${\to}$ $B_{c}{\rho}$
  decay is defined as below \cite{prd66},
   \begin{equation}
  {\cal A}({\Upsilon}(1S){\to}B_{c}{\rho})\ =\
  {\cal A}_{L}({\epsilon}_{1}^{{\parallel}},{\epsilon}_{3}^{{\parallel}})
 +{\cal A}_{N}({\epsilon}_{1}^{{\perp}},{\epsilon}_{3}^{{\perp}})
 +i\,{\cal A}_{T}\,{\varepsilon}_{{\mu}{\nu}{\alpha}{\beta}}\,
  {\epsilon}_{1}^{{\mu}}\,{\epsilon}_{3}^{{\nu}}\,
   p_{1}^{\alpha}\,p_{3}^{\beta}
   \label{eq:amp01},
   \end{equation}
  which is conventionally written as the helicity amplitudes \cite{prd66},
   \begin{equation}
  {\cal A}_{0}\ =\ -C_{\cal A}\,\sum\limits_{i}
  {\cal A}_{L}^{i}({\epsilon}_{1}^{{\parallel}},{\epsilon}_{3}^{{\parallel}})
   \label{eq:amp02},
   \end{equation}
   \begin{equation}
  {\cal A}_{\parallel}\ =\ \sqrt{2}\,C_{\cal A} \sum\limits_{i}
  {\cal A}_{N}^{i}({\epsilon}_{1}^{{\perp}},{\epsilon}_{3}^{{\perp}})
   \label{eq:amp03},
   \end{equation}
   \begin{equation}
  {\cal A}_{\perp}\ =\ \sqrt{2}\,C_{\cal A}\,m_{1}\,p \sum\limits_{i}
  {\cal A}_{T}^{i}
   \label{eq:amp04},
   \end{equation}
   \begin{equation}
  C_{\cal A}\ =\ i\frac{G_{F}}{\sqrt{2}}\,\frac{C_{F}}{N}\,
  {\pi}\, f_{\Upsilon}\,f_{B_{c}}\, V_{cb} V_{ud}^{\ast}
   \label{eq:amp05},
   \end{equation}
  where $C_{F}$ $=$ $4/3$ and the color number $N$ $=$ $3$;
  the superscript $i$ on ${\cal A}_{L,N,T}^{i}$ corresponds
  to the indices of Fig.\ref{feynman}. The explicit expressions
  of building blocks ${\cal A}_{L,N,T}^{i}$ are collected in
  Appendix \ref{blocks}.

  \section{Numerical results and discussion}
  \label{sec03}

  In the rest frame of the ${\Upsilon}(1S)$ meson,
  branching ratio (${\cal B}r$), polarization fractions
  ($f_{0,{\parallel},{\perp}}$) and relative phase between
  helicity amplitudes (${\phi}_{{\parallel},{\perp}}$)
  for the ${\Upsilon}(1S)$ ${\to}$
  $B_{c}{\rho}$ weak decay are defined as
   \begin{equation}
  {\cal B}r\ =\ \frac{1}{12{\pi}}\,
   \frac{p}{m_{{\Upsilon}}^{2}{\Gamma}_{{\Upsilon}}}\, \Big\{
  {\vert}{\cal A}_{0}{\vert}^{2}+{\vert}{\cal A}_{\parallel}{\vert}^{2}
 +{\vert}{\cal A}_{\perp}{\vert}^{2} \Big\}
   \label{br},
   \end{equation}
   \begin{equation}
  f_{0,{\parallel},{\perp}}\ =\
   \frac{ {\vert}{\cal A}_{0,{\parallel},{\perp}}{\vert}^{2} }{
  {\vert}{\cal A}_{0}{\vert}^{2}+{\vert}{\cal A}_{\parallel}{\vert}^{2}
 +{\vert}{\cal A}_{\perp}{\vert}^{2} }
   \label{f0},
   \end{equation}
   \begin{equation}
  {\phi}_{{\parallel},{\perp}}\ =\ {\arg} (
   {\cal A}_{{\parallel},{\perp}} / {\cal A}_{0} )
   \label{phi},
   \end{equation}
  where mass $m_{{\Upsilon}}$ $=$ $9460.30{\pm}0.26$ MeV
  and decay width ${\Gamma}_{\Upsilon}$ $=$
  $54.02{\pm}1.25$ keV \cite{pdg}.

  The values of other input parameters are listed as follows.
  If not specified explicitly, we will take their
  central values as default inputs.

  (1) Wolfenstein parameters \cite{pdg}:
    $A$ $=$ $0.814^{+0.023}_{-0.024}$ and
    ${\lambda}$ $=$ $0.22537{\pm}0.00061$.

  (2) Masses of quarks \cite{pdg}:
    $m_{c}$ $=$ $1.67{\pm}0.07$ GeV and
    $m_{b}$ $=$ $4.78{\pm}0.06$ GeV.

  (3) Gegenbauer moments\footnotemark[1]
   \footnotetext[1]{$a^{\parallel}_{0}$ $=$ $1$ is due to the normalization
   condition ${\int}_{0}^{1}{\phi}_{\rho}^{v}(x)dx$ $=$ $1$. More discussion
   on the ${\rho}$ wave functions and Gegenbauer moments $a^{\parallel}_{2}$
   can be found in the recent references, such as Ref.\cite{wu3}.}
    $a^{\parallel}_{0}$ $=$ $1$ and
    $a^{\parallel}_{2}$ $=$ $0.15{\pm}0.07$ for twist-2
    distribution amplitudes of the ${\rho}$ meson \cite{jhep0605}.

  (4) Decay constants:
    $f_{{\Upsilon}}$ $=$ $(676.4{\pm}10.7)$ MeV \cite{fbb},
    $f_{B_{c}}$ $=$ $489{\pm}5$ MeV \cite{fbc},
    $f_{\rho}$ $=$ $216{\pm}3$ MeV \cite{jhep0605}.

  Our numerical results are presented as follows:
  \begin{equation}
 {\cal B}r\, =\, (8.34^{+0.47+1.35+0.40+1.44}_{-0.69-0.88-0.40-1.26}){\times}10^{-9}
  \label{num:br},
  \end{equation}
  \begin{equation}
  f_{0}\, =\, (82.2^{+ 0.0+ 1.1+ 0.0}_{- 0.7- 1.3- 0.0})\%
  \label{num:br-f0},
  \end{equation}
  \begin{equation}
  f_{{\parallel}}\, =\, (15.0^{+ 0.6+ 1.0+ 0.0}_{- 0.0- 0.8- 0.0})\%
  \label{num:br-fpara},
  \end{equation}
  \begin{equation}
  f_{\perp}\, =\, (2.8^{+ 0.1+ 0.3+ 0.0}_{- 0.0- 0.3- 0.0})\%
  \label{num:br-fperp},
  \end{equation}
  \begin{equation}
  {\phi}_{{\parallel}}\, {\simeq}\, 0, \qquad
  {\phi}_{{\perp}}\, {\simeq}\, {\pi}
  \label{num:phase-fperp},
  \end{equation}
  where the first uncertainty comes from the choice of the typical
  scale $(1{\pm}0.1)t_{i}$, and the expression $t_{i}$ is
  given in Eq.(\ref{tab}) and Eq.(\ref{tcd});
  the second uncertainty is from masses $m_{b}$ and $m_{c}$;
  the third uncertainty is from hadronic parameters including
  decay constants and Gegenbauer moments; and the fourth
  uncertainty of branching ratio comes from the CKM parameters.
  The following are some comments.

  (1)
  Branching ratio for the ${\Upsilon}(1S)$ ${\to}$ $B_{c}{\rho}$
  decay with the pQCD approach is different from previous
  estimation \cite{ijma14,adv2013} with the NF approximation.
  Many factors lead to these differences.
  For example, as it is showed in Ref. \cite{adv2013}, the
  values of form factors for ${\Upsilon}(1S)$ ${\to}$ $B_{c}$
  transition are very sensitive to the choice of wave functions.
  In addition, form factors written as the convolution integral
  of wave functions in Ref. \cite{adv2013} are usually
  enhanced by one-gluon-exchange scattering amplitudes with
  the pQCD approach.
  These discrepancy deserve much dedicated study and should be
  carefully tested by the future experiments.

  (2)
  Branching ratio for the ${\Upsilon}(1S)$ ${\to}$ $B_{c}{\rho}$
  decay can reach up to ${\cal O}(10^{-9})$, which
  might be measurable at the running LHC and forthcoming SuperKEKB.
  For example, the ${\Upsilon}(1S)$ production cross section in
  p-Pb collision is about a few ${\mu}b$ at LHCb \cite{jhep1407}
  and ALICE \cite{plb740}. Over $10^{12}$ ${\Upsilon}(1S)$
  data samples per $ab^{-1}$ data collected at LHCb and ALICE
  are in principle available, corresponding to a few thousands
  of the ${\Upsilon}(1S)$ ${\to}$ $B_{c}{\rho}$ events.

  (3)
  There is a hierarchical pattern among the
  longitudinal $f_{0}$, parallel $f_{{\parallel}}$,
  and perpendicular $f_{{\perp}}$ polarization fractions, i.e.,
  \begin{equation}
  f_{0}:f_{{\parallel}}:f_{{\perp}}\ {\simeq}\
  1: \frac{p}{\sqrt{2}m_{{\Upsilon}(1S)}} :\frac{p^{2}}{2m_{{\Upsilon}(1S)}^{2}}
  \label{fff},
  \end{equation}
  where $p$ is the common momentum of final state in the
  rest frame of the ${\Upsilon}(1S)$ meson.
  The relation Eq.(\ref{fff}) is basically agree
  with previous estimation \cite{adv2013}.
  It means that the contributions to branching ratio for
  the ${\Upsilon}(1S)$ ${\to}$ $B_{c}{\rho}$ decay mainly
  come from the longitudinal polarization fractions,
  because of $f_{0}$ $>$ $f_{{\parallel}}$ $>$ $f_{{\perp}}$.

  (4)
  The relative phase ${\phi}_{{\parallel}}$ is close to zero.
  The reason is that the factorizable contributions
  from diagrams Fig.\ref{feynman}(a,b) is real and
  proportional to the large coefficient $a_{1}$, while
  the nonfactorizable contributions from diagrams
  Fig.\ref{feynman}(c,d) is suppressed by the color
  factor and proportional to the small Wilson
  coefficient $C_{2}$, and the strong phases arise
  only from the nonfactorizable contributions,
  which is consistent with the prediction of the QCD
  factorization approach \cite{qcdf1,qcdf2}
  where the strong phase arising from nonfactorizable
  contributions is suppressed by color and
  ${\alpha}_{s}$ for the $a_{1}$-dominated
  processes. The relative phases, if they could be
  determined experimentally, will improve
  our understanding on the strong interactions.

  \section{Summary}
  \label{sec04}
  The ${\Upsilon}(1S)$ weak decay is allowable within the
  standard model. In this paper,
  the ${\Upsilon}(1S)$ ${\to}$ $B_{c}{\rho}$ weak decays
  are studied with the pQCD approach.
  It is found that with the nonrelativistic wave functions
  for ${\Upsilon}(1S)$ and $B_{c}$ mesons, the longitudinal
  polarization fraction is the largest one, and branching
  ratios for the ${\Upsilon}(1S)$ ${\to}$ $B_{c}{\rho}$
  decay can reach up to ${\cal O}(10^{-9})$, which might
  be detectable at the future experiments.

  \section*{Acknowledgments}
  We thank Professor Dongsheng Du (IHEP@CAS),
  Professor Caidian L\"{u} (IHEP@CAS) and Professor
  Yadong Yang (CCNU) for helpful discussion.
  The work is supported by the National Natural Science Foundation
  of China (Grant Nos. 11547014, 11475055, U1332103 and 11275057).

  \begin{appendix}
  \section{Building blocks of decay amplitudes}
  \label{blocks}
  For the sake of simplicity, the amplitude for the ${\Upsilon}(1S)$
  ${\to}$ $B_{c}{\rho}$ decay, Eq.(\ref{eq:amp01}), is decomposed
  into building blocks ${\cal A}^{i}_{L,N,T}$, where the superscript
  $i$ corresponds to the indices of Fig.\ref{feynman}.
  With the pQCD master formula Eq.(\ref{hadronic}),
  the explicit expressions of ${\cal A}^{i}_{L,N,T}$
  are written as follows:
   \begin{eqnarray}
  {\cal A}_{L}^{a} &=&
  {\int}_{0}^{1}dx_{1}  {\int}_{0}^{1}dx_{2}
  {\int}_{0}^{\infty}b_{1}db_{1}
  {\int}_{0}^{\infty}b_{2}db_{2}\,
  {\phi}_{\Upsilon}^{v}(x_{1})
   \nonumber \\ & &
  {\phi}_{B_{c}}(x_{2})\, E_{f}(t_{a})\,
  {\alpha}_{s}(t_{a})\, a_{1}(t_{a})\,
  H_{f}({\alpha}_{e},{\beta}_{a},b_{1},b_{2})
   \nonumber \\ & &
   \Big\{ m_{1}^{2}\,s+m_{2}\,m_{b}\,u-
   (4\,m_{1}^{2}\,p^{2}+m_{2}^{2}\,u)\,\bar{x}_{2} \Big\}
   \label{amp:al}, \\
  {\cal A}_{N}^{a} &=& m_{1}\, m_{3}
  {\int}_{0}^{1}dx_{1}  {\int}_{0}^{1}dx_{2}
  {\int}_{0}^{\infty}b_{1}db_{1}
  {\int}_{0}^{\infty}b_{2}db_{2}\,
  {\phi}_{\Upsilon}^{V}(x_{1})
   \nonumber \\ & &
  {\phi}_{B_{c}}(x_{2})\, E_{f}(t_{a})\,
  {\alpha}_{s}(t_{a})\,a_{1}(t_{a})\,
  H_{f}({\alpha}_{e},{\beta}_{a},b_{1},b_{2})\,
   \nonumber \\ & &
   \Big\{  2\,m_{2}^{2}\,\bar{x}_{2} -2\,m_{2}\,m_{b} -t \Big\}
   \label{amp:an}, \\
  {\cal A}_{T}^{a} &=& 2\, m_{1}\,m_{3}\,
  {\int}_{0}^{1}dx_{1}  {\int}_{0}^{1}dx_{2}
  {\int}_{0}^{\infty}b_{1}db_{1}
  {\int}_{0}^{\infty}b_{2}db_{2}\,
  {\phi}_{\Upsilon}^{V}(x_{1})
   \nonumber \\ & &
  {\phi}_{B_{c}}(x_{2})\,
   E_{f}(t_{a})\, {\alpha}_{s}(t_{a})\,a_{1}(t_{a})\,
   H_{f}({\alpha}_{e},{\beta}_{a},b_{1},b_{2})
   \label{amp:at},
   \end{eqnarray}
   \begin{eqnarray}
  {\cal A}_{L}^{b} &=&
  {\int}_{0}^{1}dx_{1}  {\int}_{0}^{1}dx_{2}
  {\int}_{0}^{\infty}b_{1}db_{1}
  {\int}_{0}^{\infty}b_{2}db_{2}
   \nonumber \\ & &
  {\phi}_{B_{c}}(x_{2})\,E_{f}(t_{b})\,
  {\alpha}_{s}(t_{b})\,a_{1}(t_{b})
   H_{f}({\alpha}_{e},{\beta}_{b},b_{2},b_{1})
   \nonumber \\ & &
   \Big\{ {\phi}_{\Upsilon}^{v}(x_{1})\, \Big[m_{1}^{2}\,
   (s-4\,p^{2})\,\bar{x}_{1}+2\,m_{2}\,m_{c}\,u-m_{2}^{2}\,u \Big]
   \nonumber \\ & &
   + {\phi}_{\Upsilon}^{t}(x_{1})\, m_{1}\,\Big[
   s\,(2\,m_{2}-m_{c})-2\,m_{2}\,u\,\bar{x}_{1} \Big] \Big\}
   \label{amp:bl}, \\
  {\cal A}_{N}^{b} &=& m_{3}
  {\int}_{0}^{1}dx_{1}  {\int}_{0}^{1}dx_{2}
  {\int}_{0}^{\infty}b_{1}db_{1}
  {\int}_{0}^{\infty}b_{2}db_{2}
   \nonumber \\ & &
  {\phi}_{B_{c}}(x_{2})\,E_{f}(t_{b})\,
  {\alpha}_{s}(t_{b})\, a_{1}(t_{b})\,
  H_{f}({\alpha}_{e},{\beta}_{b},b_{2},b_{1})
   \nonumber \\ & &
   \Big\{ {\phi}_{\Upsilon}^{V}(x_{1})\,m_{1}\,
   \Big[ 2\,m_{2}^{2} -4\,m_{2}\,m_{c} -t\,\bar{x}_{1} \Big]
   \nonumber \\ & &
   + {\phi}_{\Upsilon}^{T}(x_{1})\, \Big[
   t\,(m_{c}-2\,m_{2})+4\,m_{1}^{2}\,m_{2}\,\bar{x}_{1} \Big] \Big\}
   \label{amp:bn}, \\
  {\cal A}_{T}^{b} &=& -2\, m_{3}\,
  {\int}_{0}^{1}dx_{1}  {\int}_{0}^{1}dx_{2}
  {\int}_{0}^{\infty}b_{1}db_{1}
  {\int}_{0}^{\infty}b_{2}db_{2}
   \nonumber \\ & &
  {\phi}_{B_{c}}(x_{2})\, E_{f}(t_{b})\,
  {\alpha}_{s}(t_{b})\, a_{1}(t_{b})\,
  H_{f}({\alpha}_{e},{\beta}_{b},b_{2},b_{1})
  \nonumber \\ & &
  \Big\{ {\phi}_{\Upsilon}^{V}(x_{1})\,m_{1}\,\bar{x}_{1}
  + {\phi}_{\Upsilon}^{T}(x_{1})\, (m_{c}-2\,m_{2}) \Big\}
   \label{amp:bt},
   \end{eqnarray}
   \begin{eqnarray}
  {\cal A}_{L}^{c} &=& \frac{1}{N_{c}}
  {\int}_{0}^{1}dx_{1} {\int}_{0}^{1}dx_{2} {\int}_{0}^{1}dx_{3}
  {\int}_{0}^{\infty}db_{1}  {\int}_{0}^{\infty}b_{2}db_{2}
  {\int}_{0}^{\infty}b_{3}db_{3}
   \nonumber \\ & &
  {\phi}_{B_{c}}(x_{2})\, {\phi}_{\rho}^{v}(x_{3})\,
   E_{n}(t_{c})\, {\alpha}_{s}(t_{c})\, C_{2}(t_{c})\,
   H_{n}({\alpha}_{e},{\beta}_{c},b_{2},b_{3})
   \nonumber \\ & &
  {\delta}(b_{1}-b_{2})\,
   \Big\{ {\phi}_{\Upsilon}^{v}(x_{1})\,u\,
   \Big[ t\,x_{1}-2\,m_{2}^{2}\,x_{2}-s\,\bar{x}_{3} \Big]
   \nonumber \\ & &
   + {\phi}_{\Upsilon}^{t}(x_{1})\, m_{1}\,m_{2}\,
   \Big[ s\,x_{2}+2\,m_{3}^{2}\,\bar{x}_{3}-u\,x_{1}\Big] \Big\}
   \label{amp:cl}, \\
  {\cal A}_{N}^{c} &=& \frac{ m_{3} }{N_{c}}
  {\int}_{0}^{1}dx_{1} {\int}_{0}^{1}dx_{2} {\int}_{0}^{1}dx_{3}
  {\int}_{0}^{\infty}db_{1}  {\int}_{0}^{\infty}b_{2}db_{2}
  {\int}_{0}^{\infty}b_{3}db_{3}
   \nonumber \\ & &
   {\phi}_{B_{c}}(x_{2})\, E_{n}(t_{c})\, {\alpha}_{s}(t_{c})\,
   C_{2}(t_{c})\, H_{n}({\alpha}_{e},{\beta}_{c},b_{2},b_{3})
   \nonumber \\ & &
   \Big\{ {\phi}_{\Upsilon}^{V}(x_{1})\,{\phi}_{\rho}^{V}(x_{3})\,m_{1}\,
   \Big[ 2\,s\,\bar{x}_{3}+4\,m_{2}^{2}\,x_{2}\,-2\,t\,x_{1} \Big]
   \nonumber \\ & &
   + {\phi}_{\Upsilon}^{T}(x_{1})\,{\phi}_{\rho}^{V}(x_{3})\,m_{2}\,
   \Big[ 2\,m_{1}^{2}\,x_{1} -t\,x_{2}-u\,\bar{x}_{3} \Big]
   \nonumber \\ & &
   + {\phi}_{\Upsilon}^{T}(x_{1})\,{\phi}_{\rho}^{A}(x_{3})\,
   2\,m_{1}\,m_{2}\,p\,( x_{2}-\bar{x}_{3} ) \Big\}\,
   {\delta}(b_{1}-b_{2})
   \label{amp:cn}, \\
  {\cal A}_{T}^{c} &=& \frac{ m_{3} }{N_{c}\,p}
  {\int}_{0}^{1}dx_{1} {\int}_{0}^{1}dx_{2} {\int}_{0}^{1}dx_{3}
  {\int}_{0}^{\infty}db_{1}  {\int}_{0}^{\infty}b_{2}db_{2}
  {\int}_{0}^{\infty}b_{3}db_{3}\,
   \nonumber \\ & &
   {\phi}_{B_{c}}(x_{2})\, E_{n}(t_{c})\, {\alpha}_{s}(t_{c})\,
   C_{2}(t_{c})\, H_{n}({\alpha}_{e},{\beta}_{c},b_{2},b_{3})
   \nonumber \\ & &
   \Big\{ {\phi}_{\Upsilon}^{V}(x_{1})\,{\phi}_{\rho}^{A}(x_{3})\,
   \Big[ 2\,s\,\bar{x}_{3}+4\,m_{2}^{2}\,x_{2}\,-2\,t\,x_{1} \Big]
   \nonumber \\ & &
   + {\phi}_{\Upsilon}^{T}(x_{1})\, {\phi}_{\rho}^{A}(x_{3})\,r_{2}\,
   \Big[ 2\,m_{1}^{2}\,x_{1} -t\,x_{2}-u\,\bar{x}_{3} \Big]
   \nonumber \\ & &
   +2\,m_{2}\,p\, {\phi}_{\Upsilon}^{T}(x_{1})\, {\phi}_{\rho}^{V}(x_{3})\,
   (x_{2}-\bar{x}_{3}) \Big\}\, {\delta}(b_{1}-b_{2})
   \label{amp:ct},
   \end{eqnarray}
   \begin{eqnarray}
  {\cal A}_{L}^{d} &=& \frac{1}{N_{c}}
  {\int}_{0}^{1}dx_{1} {\int}_{0}^{1}dx_{2} {\int}_{0}^{1}dx_{3}
  {\int}_{0}^{\infty}db_{1}  {\int}_{0}^{\infty}b_{2}db_{2}
  {\int}_{0}^{\infty}b_{3}db_{3}
   \nonumber \\ & &
  {\phi}_{B_{c}}(x_{2})\, {\phi}_{\rho}^{v}(x_{3})\,
   E_{n}(t_{d})\, {\alpha}_{s}(t_{d})\, C_{2}(t_{d})\,
   H_{n}({\alpha}_{e},{\beta}_{d},b_{2},b_{3})
   \nonumber \\ & &
  {\delta}(b_{1}-b_{2})\,
   \Big\{ {\phi}_{\Upsilon}^{t}(x_{1})\, m_{1}\,m_{2}\,
   \Big[ s\,x_{2}+2\,m_{3}^{2}\,x_{3}-u\,x_{1}\Big]
   \nonumber \\ & &
 +{\phi}_{\Upsilon}^{v}(x_{1})\,4\,m_{1}^{2}\,p^{2}\,
   (x_{3}-x_{2}) \Big\}
   \label{amp:dl}, \\
  {\cal A}_{N}^{d} &=& \frac{ m_{3} }{N_{c}}
  {\int}_{0}^{1}dx_{1} {\int}_{0}^{1}dx_{2} {\int}_{0}^{1}dx_{3}
  {\int}_{0}^{\infty}db_{1}  {\int}_{0}^{\infty}b_{2}db_{2}
  {\int}_{0}^{\infty}b_{3}db_{3}
   \nonumber \\ & &
   {\phi}_{\Upsilon}^{T}(x_{1})\, {\phi}_{B_{c}}(x_{2})\,
   E_{n}(t_{d})\, {\alpha}_{s}(t_{d})\,
   C_{2}(t_{d})\, H_{n}({\alpha}_{e},{\beta}_{d},b_{2},b_{3})
   \nonumber \\ & &
   {\delta}(b_{1}-b_{2})\,
   \Big\{ {\phi}_{\rho}^{V}(x_{3})\,m_{2}\,
   \Big[ 2\,m_{1}^{2}\,x_{1}-t\,x_{2}\,-u\,x_{3} \Big]
   \nonumber \\ & &
   + 2\,m_{1}\,m_{2}\,p\,{\phi}_{\rho}^{A}(x_{3})\,(x_{2}-x_{3}) \Big\}
   \label{amp:dn}, \\
  {\cal A}_{T}^{d} &=& \frac{ m_{3} }{N_{c}\,p}
  {\int}_{0}^{1}dx_{1} {\int}_{0}^{1}dx_{2} {\int}_{0}^{1}dx_{3}
  {\int}_{0}^{\infty}db_{1}  {\int}_{0}^{\infty}b_{2}db_{2}
  {\int}_{0}^{\infty}b_{3}db_{3}
   \nonumber \\ & &
   {\phi}_{\Upsilon}^{T}(x_{1})\, {\phi}_{B_{c}}(x_{2})\,
   E_{n}(t_{d})\, {\alpha}_{s}(t_{d})\,
   C_{2}(t_{d})\, H_{n}({\alpha}_{e},{\beta}_{d},b_{2},b_{3})
   \nonumber \\ & &
   {\delta}(b_{1}-b_{2})\,
   \Big\{ {\phi}_{\rho}^{A}(x_{3})\,r_{2}\,
   \Big[ 2\,m_{1}^{2}\,x_{1} -t\,x_{2}-u\,x_{3} \Big]
   \nonumber \\ & &
   +2\,m_{2}\,p\,{\phi}_{\rho}^{V}(x_{3})\,(x_{2}-x_{3}) \Big\}
   \label{amp:dt},
   \end{eqnarray}
  where
  $\bar{x}_{i}$ $=$ $1$ $-$ $x_{i}$;
  variable $x_{i}$ and $b_{i}$ are the longitudinal momentum
  fraction and the conjugate variable of the transverse
  momentum $k_{i{\perp}}$ of the valence quark, respectively;
  ${\alpha}_{s}$ is the QCD coupling;
  $a_{1}$ $=$ $C_{1}$ $+$ $C_{2}/N$;
  $C_{1,2}$ are the Wilson coefficients.

  The function $H_{f,n}$ and Sudakov factor $E_{f,n}$ are defined
  as follows, where the subscripts $f$ and $n$ correspond to
  factorizable and nonfactorizable topologies, respectively.
   \begin{equation}
   H_{f}({\alpha}_{e},{\beta},b_{i},b_{j}) \ =\
   K_{0}(\sqrt{-{\alpha}_{e}}b_{i})
   \Big\{ {\theta}(b_{i}-b_{j})
   K_{0}(\sqrt{-{\beta}}b_{i})
   I_{0}(\sqrt{-{\beta}}b_{j})
   + (b_{i}{\leftrightarrow}b_{j}) \Big\}
   \label{hab},
   \end{equation}
   \begin{eqnarray}
   H_{n}({\alpha}_{e},{\beta},b_{2},b_{3}) &=&
   \Big\{ {\theta}(-{\beta}) K_{0}(\sqrt{-{\beta}}b_{3})
  +\frac{{\pi}}{2} {\theta}({\beta}) \Big[
   iJ_{0}(\sqrt{{\beta}}b_{3})
   -Y_{0}(\sqrt{{\beta}}b_{3}) \Big] \Big\}
   \nonumber \\ &{\times}&
   \Big\{ {\theta}(b_{2}-b_{3})
   K_{0}(\sqrt{-{\alpha}_{e}}b_{2})
   I_{0}(\sqrt{-{\alpha}_{e}}b_{3})
   + (b_{2}{\leftrightarrow}b_{3}) \Big\}
   \label{hcd},
   \end{eqnarray}
   \begin{equation}
   E_{f}(w)\ =\ {\exp}\{ -S_{\Upsilon}(w)-S_{B_{c}}(w) \}
   \label{sudakov-f},
   \end{equation}
   \begin{equation}
   E_{n}(w)\ =\ {\exp}\{ -S_{\Upsilon}(w)-S_{B_{c}}(w)-S_{\rho}(w) \}
   \label{sudakov-n},
   \end{equation}
   \begin{equation}
   S_{\Upsilon}(w)\ =\
   s(x_{1},p_{1}^{+},1/b_{1})
  +2{\int}_{1/b_{1}}^{w}\frac{d{\mu}}{\mu}{\gamma}_{q}
   \label{sudakov-bb},
   \end{equation}
   \begin{equation}
   S_{B_{c}}(w)\ =\
   s(x_{2},p_{2}^{+},1/b_{2})
  +2{\int}_{1/b_{2}}^{w}\frac{d{\mu}}{\mu}{\gamma}_{q}
   \label{sudakov-bc},
   \end{equation}
   \begin{equation}
   S_{\rho}(w)\ =\
   s(x_{3},p_{3}^{+},1/b_{3})
  +s(\bar{x}_{3},p_{3}^{+},1/b_{3})
  +2{\int}_{1/b_{3}}^{w}\frac{d{\mu}}{\mu}{\gamma}_{q}
   \label{sudakov-ds},
   \end{equation}
  where $J_{0}$ and $Y_{0}$ ($I_{0}$ and $K_{0}$) are the
  (modified) Bessel function of the first and second kind,
  respectively;
  ${\gamma}_{q}$ $=$ $-{\alpha}_{s}/{\pi}$ is the
  quark anomalous dimension; the expression of $s(x,Q,1/b)$
  can be found in the appendix of Ref.\cite{pqcd1};
  ${\alpha}_{e}$ is the gluon virtuality;
  the subscript of the quark virtuality ${\beta}_{i}$
  corresponds to the indices of Fig.\ref{feynman}.
  The definitions of the particle virtuality and typical
  scale $t_{i}$ are listed as follows:
   \begin{eqnarray}
  {\alpha}_{e} &=& \bar{x}_{1}^{2}m_{1}^{2}
                +  \bar{x}_{2}^{2}m_{2}^{2}
                -  \bar{x}_{1}\bar{x}_{2}t
   \label{gluon-q2-e}, \\
  {\beta}_{a} &=& m_{1}^{2} - m_{b}^{2}
               +  \bar{x}_{2}^{2}m_{2}^{2}
               -  \bar{x}_{2}t
   \label{beta-fa}, \\
  {\beta}_{b} &=& m_{2}^{2} - m_{c}^{2}
               +  \bar{x}_{1}^{2}m_{1}^{2}
               -  \bar{x}_{1}t
   \label{beta-fb}, \\
  {\beta}_{c} &=& x_{1}^{2}m_{1}^{2}
               +  x_{2}^{2}m_{2}^{2}
               +  \bar{x}_{3}^{2}m_{3}^{2}
   \nonumber \\ &-&
                  x_{1}x_{2}t
               -  x_{1}\bar{x}_{3}u
               +  x_{2}\bar{x}_{3}s
   \label{beta-fc}, \\
  {\beta}_{d} &=& x_{1}^{2}m_{1}^{2}
               +  x_{2}^{2}m_{2}^{2}
               +  x_{3}^{2}m_{3}^{2}
    \nonumber \\ &-&
                  x_{1}x_{2}t
               -  x_{1}x_{3}u
               +  x_{2}x_{3}s
   \label{beta-fd}, \\
   t_{a(b)} &=& {\max}(\sqrt{-{\alpha}_{e}},\sqrt{-{\beta}_{a(b)}},1/b_{1},1/b_{2})
   \label{tab}, \\
   t_{c(d)} &=& {\max}(\sqrt{-{\alpha}_{e}},\sqrt{{\vert}{\beta}_{c(d)}{\vert}},1/b_{2},1/b_{3})
   \label{tcd}.
   \end{eqnarray}

  \end{appendix}

  
  \end{document}